\documentclass[11pt,prl,aps,amsfonts,amsmath,amssymb,nofootinbib,superscriptaddress]{revtex4-2}
\usepackage{mathtools}
\usepackage{upgreek}
\usepackage{lineno}
\usepackage{graphicx}
\usepackage{bm}
\usepackage{hyperref}
\usepackage{braket}
\newcommand{\RNum}[1]{\uppercase\expandafter{\romannumeral #1\relax}}
\usepackage{natbib}
\usepackage{float}
\usepackage{xcolor}
\usepackage{svg}
\restylefloat{table}

\begin{document}
\title{Topological phase transition to a hidden charge density wave liquid
}
\author{Joshua S.H. Lee}
\thanks{These authors contributed equally.}
\affiliation{Department of Physics and Astronomy, University of California Los Angeles, Los Angeles, CA 90095, USA}

\author{Thomas M. Sutter}
\thanks{These authors contributed equally.}
\affiliation{Department of Physics and Astronomy, University of California Los Angeles, Los Angeles, CA 90095, USA}

\author{Goran Karapetrov}
\affiliation{Department of Physics, Drexel University, Philadelphia, Pennsylvania 19104, USA}

\author{Pietro Musumeci}
\affiliation{Department of Physics and Astronomy, University of California Los Angeles, Los Angeles, CA 90095, USA}

\author{Anshul Kogar}
\email{anshulkogar@physics.ucla.edu}
\affiliation{Department of Physics and Astronomy, University of California Los Angeles, Los Angeles, CA 90095, USA}

\date{\today}

\begin{abstract}
    \textbf{Charge density waves (CDWs), electronic crystals that form within a host solid, have long been speculated to melt into a spatially textured electronic liquid~\cite{dai1991weak, dai1992hexaticCdwNbSubt}. Though they have not been previously detected, liquid CDWs may nonetheless be fundamental to the phase diagrams of many correlated electron systems, including high temperature superconductors and quantum Hall states~\cite{hayden2024charge, subires2024frustrated, kivelson1998electronic, delacretaz2017theory, taraphder2011preformed, snow2003quantum, ciftja2004liquid, wexler2002liquidcrystalline, fradkin1999liquid, nie2014quenched, milward2005electronically}. In one of the most promising candidate materials capable of hosting a liquid CDW, 1$T$-TaS$_2$, a structural phase transition impedes its observation. Here, by irradiating the material with a femtosecond light pulse, we circumvent the structural phase transition to reveal how topological defect dynamics govern the otherwise invisible CDW correlations. Upon photoexcitation, the CDW diffraction peaks broaden azimuthally, initially revealing a hexatic state. At higher temperatures, photoexcitation completely destroys translational and orientational order and only a ring of diffuse scattering is observed, a key signature of a liquid CDW. Our work provides compelling evidence for a defect-unbinding transition to a CDW liquid and presents a protocol for uncovering states that are hidden by other transitions in thermal equilibrium.}
\end{abstract}

\begin{titlepage}
\maketitle
\end{titlepage}

An intervening phase transition can render regions of a system's thermodynamic phase space inaccessible and make challenging the experimental realization of quantum phases. Trapping gasses in a metastable state to avoid solidification was key to observing Bose-Einstein condensation, a hallmark achievement of twentieth century physics~\cite{anderson1995observation, davis1995bose}. On the other hand, liquid superconductors remain only conjectured because all known liquid metals, like mercury, form Cooper pairs below their solidification temperature~\cite{tenney2021metallic, leggett2016superconductivity}. Experiments that bypass the constraints imposed by thermodynamic equilibrium are crucial to accessing forbidden parts of a system's phase space that may host unrealized states~\cite{stojchevska2014ultrafast, kogar2020light, fausti2011light,disa2023photo,mitrano2016possible,nova2019metastable,li2019terahertz}. 

Primarily because of intervening phase transitions in candidate materials, a liquid CDW is one such state that has yet to be experimentally realized despite being speculated about for more than three decades~\cite{dai1992hexaticCdwNbSubt, dai1991weak}. A liquid CDW is distinct from an electronic Fermi liquid, as it possesses a finite order parameter amplitude with a corresponding local periodicity, but nonetheless lacks long range orientational and translational order. If observed, a liquid CDW would represent a new state of electronic matter and serve as a basis for discovering new classes of electronic liquid crystals. Liquid-like CDWs have been theoretically predicted to play a fundamental role in the anomalous normal state properties of several families of unconventional superconductors including cuprates, transition metal dichalcogenides and kagome metals as well as in the electronic properties of various quantum Hall systems~\cite{hayden2024charge,subires2024frustrated,kivelson1998electronic, delacretaz2017theory, taraphder2011preformed, snow2003quantum, ciftja2004liquid, wexler2002liquidcrystalline, fradkin1999liquid, nie2014quenched, milward2005electronically}. 

\begin{figure*}
    \centering	\includegraphics[width=1\columnwidth]{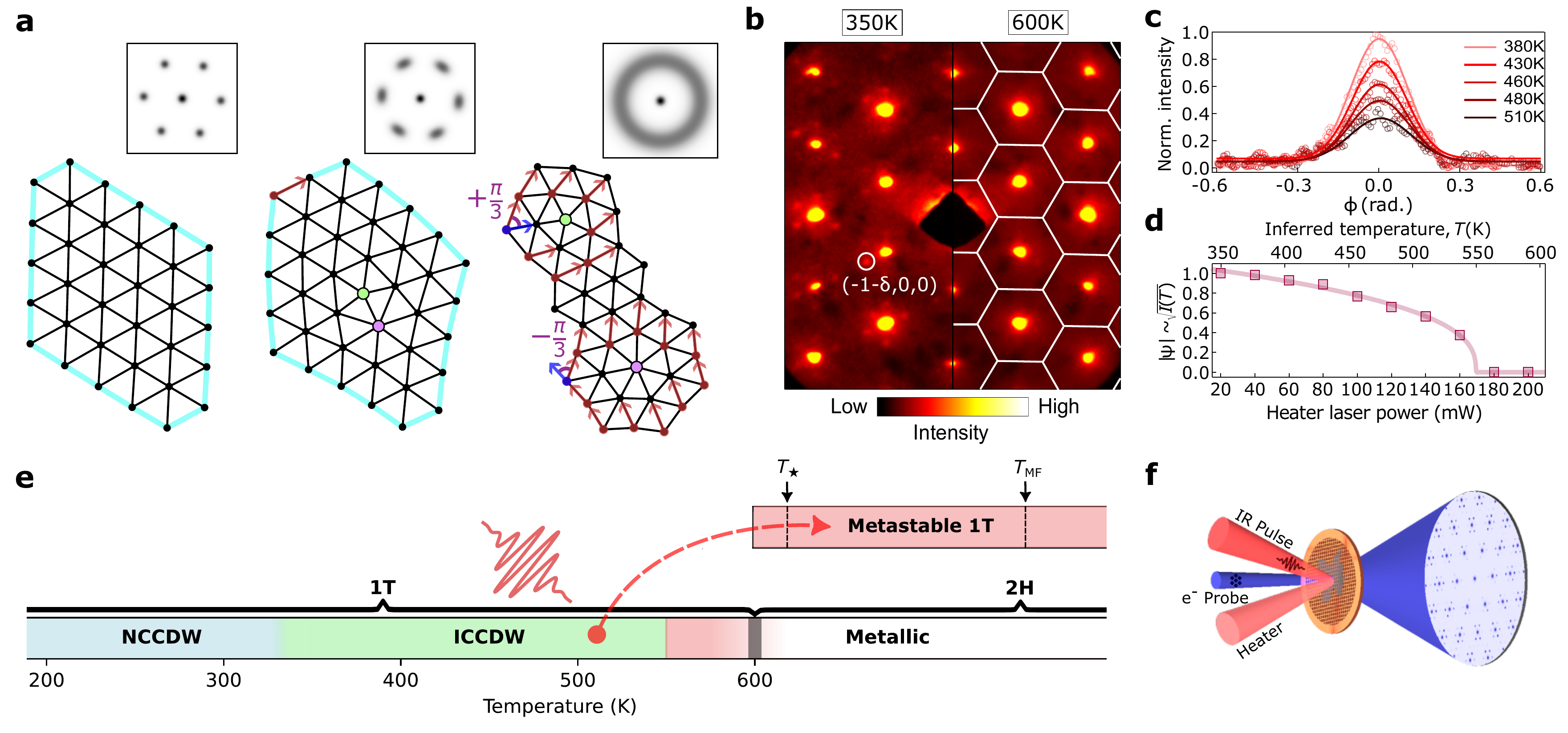}
    \caption{\textbf{Defect-mediated melting of a two-dimensional triangular lattice.} \textbf{a}, Schematic of the solid-hexatic-liquid phase transition of a triangular two-dimensional lattice. Insets show diffraction patterns corresponding to each real-space lattice. In the solid state (left), the blue path indicates that the Burgers vector is zero. In the hexatic state (middle), the red arrow indicates the Burgers vector for a path that encloses a single dislocation. In the liquid state (right), the blue arrows illustrate that the Frank angle is $+\pi/3$ $(-\pi/3)$ when transporting a vector around an isolated disclination with five (seven) nearest-neighbor lattice sites. \textbf{b}, Static electron diffraction patterns showing the IC-CDW phase at $T=350$K (left) and at $T=600$K (right). Brillouin zone outlines are shown overlaid on the $T=600$K diffraction pattern. \textbf{c}, Representative azimuthal profiles of the (-1-$\delta$,0,0) IC-CDW peak as indicated in \textbf{b} at selected temperatures. The peak width remains resolution-limited through the entire transition. \textbf{d}, Measurement of the order parameter $|\psi| \propto \sqrt{I(T)}$ of the IC-CDW-to-metallic transition, where $I(T)$ is the intensity of the (-1-$\delta$,0,0) CDW peak. Through measurements of the heater laser power needed to induce the NC-CDW-to-IC-CDW and IC-CDW-to-metallic transitions, a conversion between power and sample temperature is inferred. Solid line is a fit to $|\psi|\sim(T_c-T)^\beta$ with extracted critical exponent $\beta=0.37\pm0.04$. \textbf{e}, Phase diagram of $1T$-TaS$_2$ as a function of temperature. Above $600$K, $1T$-TaS$_2$ undergoes an irreversible, discontinuous phase transition to the $2H$ structure, which does not host an IC-CDW phase. However, application of a femtosecond light pulse can be used to circumvent this transition, allowing for transient study of a metastable $1T$ structure that is otherwise inaccessible in equilibrium. \textbf{f}, Schematic of the experimental setup showing the probe electron pulse (blue), the pump pulse (red), and the external heater laser (pink).}
    \label{fig:kthny}
\end{figure*}

\begin{figure*}[t]
    \centering	\includegraphics[width=1\columnwidth]{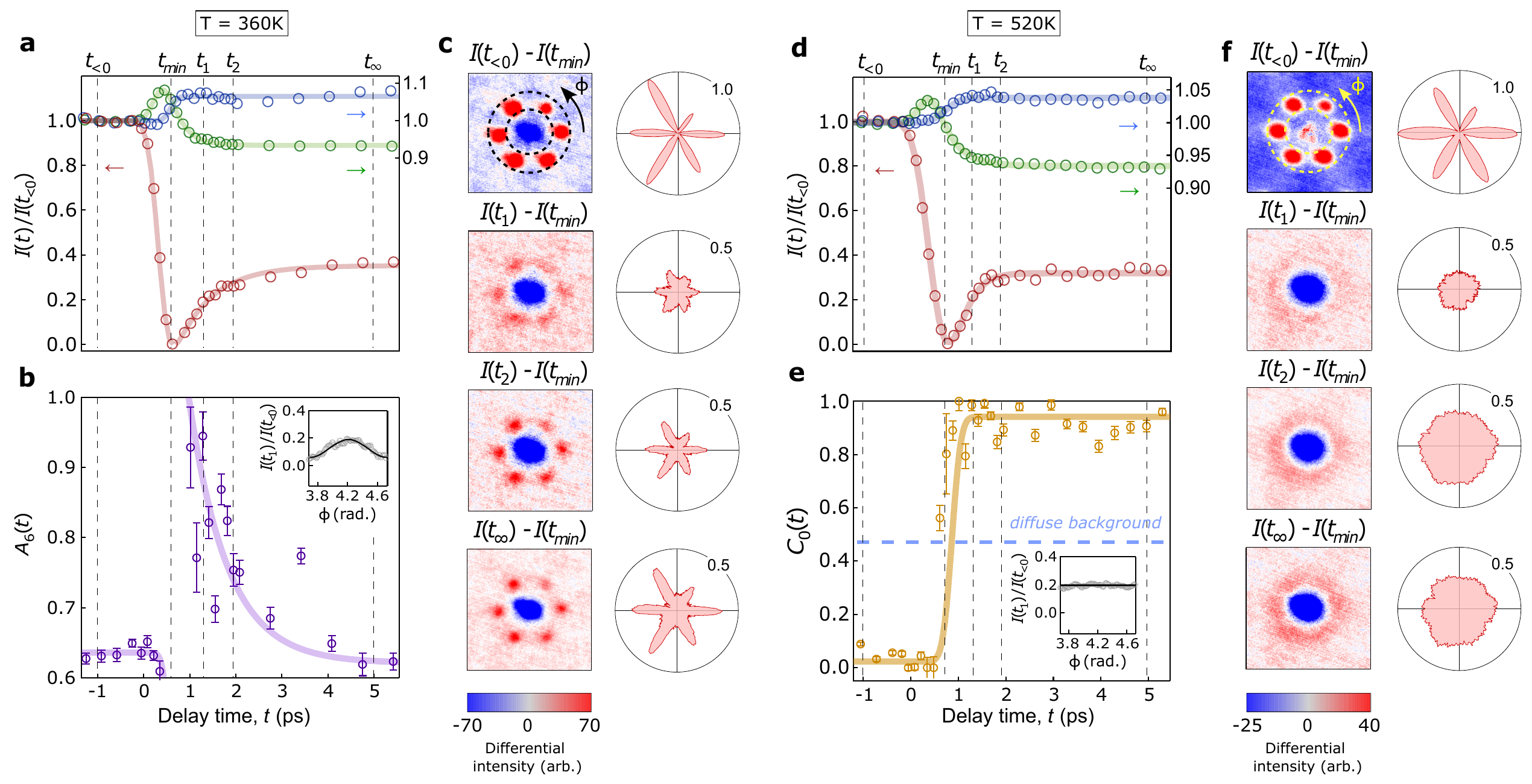}
    \caption{\textbf{Emergence of hexatic and liquid CDW states following a light-induced suppression of the equilibrium order.} \textbf{a}, Temporal dynamics of the integrated intensities of the IC-CDW peaks (red), Bragg peaks (green), and thermal diffuse background (blue) after photoexcitation, at initial temperature $T=360$K. Several representative time points are labeled: before photoexcitation $(t_{<0})$, at complete suppression of the CDW order $(t_{\text{min}})$, and at $(0.5, 1.05, 4.25)$ ps after $t_{\text{min}}$ $(t_1,t_2,t_{\infty})$. \textbf{b}, Time evolution of the fundamental 6$\phi$ Fourier coefficient $A_6(t)$ of the IC-CDW intensity from \textbf{a}. The inset shows the azimuthal IC-CDW profile of the (-1-$\delta,0,0)$ peak overlaid with $(I(t_1)/I(t_{<0}))\cos(6\phi)$, showing excellent agreement. \textbf{c}, Differential diffraction profiles at ($t_{<0}$, $t_1$, $t_2$, $t_{\infty}$) with respect to $t_{\text{min}}$ (left) and their respective azimuthal polar plots (right). \textbf{d}, Same as in \textbf{a} but at an initial temperature of $T=520$K. \textbf{e}, Time evolution of the constant offset term $C_0(t)$, representing how much the nodes have lifted in the azimuthal polar plots. $C_0(t_\infty)$ of the diffuse background is plotted in blue for reference. The inset is the same as in \textbf{b} but overlaid with a constant line, showing minimal intensity variations. \textbf{f}, Same as in \textbf{c} but at an initial temperature of $T=520$K. The azimuthal profiles at the CDW wavevector after $t=t_{\text{min}}$ show an isotropic structure distinct from the diffuse background, a signature of a liquid CDW state.}
    \label{fig:timetraces}
\end{figure*}

However, whether liquid CDWs can exist, even in principle, is controversial. Liquids possess both continuous translational and rotational symmetry, while a CDW necessarily resides within the periodic and anisotropic environment of a host crystal. Complete suppression of orientational order may be hampered due to the underlying interaction between the CDW and ionic lattice. In two dimensions (2D), however, the melting transition proceeds through the proliferation of topological defects, as described in Kosterlitz-Thouless-Halperin-Nelson-Young (KTHNY) theory, and might allow for a liquid CDW even in the presence of a crystalline background~\cite{nelson1979kthny,halperin1978kthny,kosterlitz2016review}.

Within the KTHNY framework, the solid-to-liquid phase transition evolves in two steps through an intermediate hexatic phase. Hexatics are characterized by the unbinding of dislocation pairs which results in the absence of long-range translational order but the retention of quasi-long range orientational order. Translational and rotational symmetry are ultimately restored in the liquid state through the dissociation of dislocations into individual disclinations/anti-disclinations, which are topological defects that twist the crystal lattice. In a scattering experiment, the hexatic state is marked by a preferential broadening of the diffraction peaks along the azimuthal direction. In the liquid state, the peaks completely broaden into an isotropic diffuse ring of intensity reflecting the complete loss of long-range orientational order (Fig.~\ref{fig:kthny}(a)). Solid-hexatic-liquid transitions in two dimensions have been investigated in various systems that do not, or only weakly, interact with an underlying lattice. Examples include colloidal crystals~\cite{keim2007colloidalkthny,kusner1994colloidal}, superconducting vortex lattices~\cite{troyanovski2002stm, Guillamón2009, roy2019vortexkthny}, liquid crystals~\cite{brock1986liquidcrys,cheng1988liquidcrys}, rotating Bose-Einstein condensates~\cite{sharma2024thermal}, skyrmion lattices~\cite{huang2020melting}, and Wigner crystallized free electrons on the surface of liquid helium~\cite{gallet1982wigner,knighton2018wigner}. While previous experiments on CDW materials have uncovered a hexatic state~\cite{dai1992hexaticCdwNbSubt,sung2024endotaxial,domrose2023light}, it remains unclear whether disclinations can unbind to completely remove orientational order.

One of the most promising candidate liquid CDW materials is the transition metal dichalcogenide 1$T$-TaS$_2$, where the CDW is largely two dimensional with only a weak out-of-plane coupling. Three CDWs undergo a concurrent instability, known as a triple-$\textbf{\textit{q}}$ CDW, which maintains the triangular lattice motif characteristic of KTHNY theory. Historically, a dislocation-induced hexatic state was observed in $1T$-TaS$_2$ through the substitution of niobium into the tantalum site~\cite{dai1992hexaticCdwNbSubt}. Recently, the hexatic state was observed in pristine, minimally disordered 1$T$-TaS$_2$ both upon photo-excitation and endotaxial stabilization~\cite{domrose2023light, sung2024endotaxial}; these latter studies provide a renewed impetus and a route to search for the CDW liquid.

A major obstacle in 1$T$-TaS$_2$, however, stems from an irreversible, discontinuous phase transition from the 1$T$ to 2$H$ structure that occurs at around 600K~\cite{givens1977thermal}. In a quasi-2D compound like 1$T$-TaS$_2$, the metallic to incommensurate CDW (IC-CDW) transition at $T_{\text{CDW}}(=$ 550K) occurs via a 2D-to-3D crossover~\cite{gruner2018density}. Below the mean-field transition temperature, $T_{\text{MF}}(\gg T_{CDW})$, 2D correlations begin to build up; three dimensional correlations emerge only below a temperature $T^*$, where the inter-plane correlation length exceeds the inter-plane distance. Between $T^*$ and $T_{\text{MF}}$, the KTHNY scenario should be applicable, as the system effectively functions as a stack of independent 2D CDW layers. However, both estimated temperature scales, $T^*\approx630$K and $T_{\text{MF}}\approx750$K, lie above the 1$T$-to-2$H$ polytypic phase transition temperature in a regime that cannot be accessed in thermal equilibrium. Below, we show that ultrafast electron diffraction can reveal the CDW correlations in this layer-decoupled regime by transiently maintaining the 1$T$ structure in a non-equilibrium, thermally disfavored state.


We first characterize the incommensurate CDW-to-metallic phase transition in equilibrium by obtaining static electron diffraction patterns. All diffraction measurements are performed in a transmission geometry perpendicular to the layers so that the $(\text{HK}0)$ plane is experimentally captured. Here $(\text{HKL})$ denote the Miller indices. We extract the temperature dependence of the order parameter $|\psi| \propto \sqrt{I(T)}$, where $I(T)$ is the CDW peak intensity, by increasing the power of a heater laser (Fig. \ref{fig:kthny}, Methods). Through measurements of the laser power needed to induce the nearly commensurate-to-incommensurate (NC-to-IC) CDW transition at $350\text{K}$~\cite{thomson1988nccdw} and the IC-CDW-to-metallic transition at $550\text{K}$~\cite{rossnagel2011origin}, we estimate the change in temperature of our sample induced by the external heater laser (Fig. \ref{fig:kthny}(d)). An azimuthal cut of the (-1-$\delta$, 0, 0) peak is displayed in Fig.~\ref{fig:kthny}(c), which shows the intensity of the CDW peak decreasing as the transition temperature is approached. Broadening of the peaks beyond the momentum resolution of our instrument along both the azimuthal and radial directions is not detected within the temperature range investigated (Fig. \ref{fig:kthny}(c), Supplementary Note \RNum{1} - \RNum{3}). 

In order to transiently access the layer-decoupled regime (which would correspond to $T^*<T<T_{\text{MF}}$ in thermal equilibrium), we investigate the IC-CDW-to-metallic transition upon photo-excitation with 840nm, 180fs light pulses at an incident fluence of 6mJ/cm$^2$ (Fig. \ref{fig:timetraces}). Pulsed photoexcitation allows for the sample to only be transiently heated and for the CDW to be melted non-thermally; the process by which the CDW re-crystallizes can then be investigated as a function of time. Figures \ref{fig:timetraces}(a) and (d) show the temporal evolution of the integrated intensity of the IC-CDW satellite peaks, the main Bragg peaks, and the change in the diffuse background following photoexcitation at 360K and 520K, respectively. (An estimate of the temperature before photoexcitation is obtained by comparing the intensity of the (-1-$\delta$, 0, 0) CDW peak at time $t = t_{<0}$ with its equilibrium intensity; see Supplementary Note \RNum{4}). Following photo-excitation, the IC-CDW peak intensity is suppressed within a characteristic timescale of $375$fs and reaches zero by $t=t_\text{min}\approx750$fs, values which are limited by the temporal resolution of our instrument~\cite{sutter2024ued}. Over the subsequent picosecond, the system reaches a quasi-equilibrium state. In this state, the CDW peaks only partially recover their original intensity, as the heat incurred from photo-excitation has not diffused out of the illuminated region within the measured time window. An increase in Bragg peak intensity occurs concurrently with the suppression of CDW peak intensity, a consequence of the elastic scattering sum rule. The subsequent loss of Bragg peak intensity and increase in the diffuse scattering signal arises due to lattice heating through the Debye-Waller effect and thermal occupation of phonons, respectively (Supplementary Note \RNum{8}). A number of other CDW systems have been observed to abide by a similar experimental phenomenology, albeit with different timescales~\cite{zong2019evidence,cheng2022light,gonzalez2022time,trigo2019coherent,han2012structural}.

To investigate the changes in orientational and translational order of the CDW during the recovery time frame, we examine changes to the superlattice peak profiles along the azimuthal and radial directions between $t_{\text{min}}$ and $t_{\infty}$ (Fig. \ref{fig:timetraces}(c) and (f) and Supplementary Movie). All presented differential diffraction profiles are averaged over the visible $\{ 200\}$ and $\{ 110\}$ families of CDW peaks, as well as over several time points, for improved signal-to-noise ratio (see Supplementary Note \RNum{6}). To analyze the orientational order, we perform a Fourier decomposition of the azimuthal CDW intensity profiles and study the dynamics of its coefficients after photo-excitation. The intensity can be expressed as a constant offset in addition to a Fourier series with harmonics of period six (see Supplementary Note \RNum{9}):
\begin{equation}
    I_{\text{norm}}(\phi,t)=C_0(t)+\sum\limits_{n=1}^4A_{6n}(t)\cos(6n\phi)
\end{equation}
where $\phi$ denotes the azimuthal angle around the ring shown in the top panels of Fig.~\ref{fig:timetraces}(c) and (f). Considering the momentum resolution of our instrument, harmonics higher than $n=4$ provide minimal improvements to the fit and are thus excluded. The offset term $C_0(t)$ quantifies how much the nodes lift in the polar plot visualization shown in the right columns of Fig.~\ref{fig:timetraces}(c) and (f). Importantly, the nodes can lift both due to enhanced diffuse scattering present at all momenta in addition to an isotropic scattering component at the CDW wavevector. Below, all plots of $C_0$ are therefore shown in conjunction with the background diffuse signal.

At 360K, as the CDW re-crystallizes after photo-excitation, the peaks broaden along both the azimuthal and radial directions  substantially (Fig.~\ref{fig:timetraces}(c), Fig. S6). At $t=t_1\approx 1.2$~ps, the peaks can be fit solely with the fundamental 6$\phi$ Fourier component so that $A_6\approx1$ (Fig~\ref{fig:timetraces}(b) and inset). Such broadening shows that while long-range translational order is lost, quasi-long-range orientational order is still present in the CDW superlattice. As shown in a previous work, this hexatic state occurs in the layer-decoupled regime where only 2D CDW order is present, and occurs in accordance with the unbinding of dislocation/anti-dislocation pairs~\cite{domrose2023light}. Over the next two picoseconds, the $A_6$ coefficient decreases significantly while the higher harmonic coefficients increase, observations which are indicative of peak narrowing in the azimuthal direction (Fig.~\ref{fig:timetraces}(b), Fig. S8). By $t=t_\infty$, the azimuthal and radial widths of the CDW peaks return to their values before photo-excitation ($t=t_{<0}$) and their intensities have stabilized; the system has largely restored the translationally and orientationally ordered state of the CDW superlattice. Together, these data demonstrate that the recovery of the CDW superlattice follows a non-thermal pathway through an intermediate hexatic state. 

To observe the liquid CDW state, we repeat the experiment with identical pump parameters but at an initial temperature of $520$K. While the evolution of the intensities of the CDW peaks, the main Bragg peaks, and the diffuse background is similar to those at $360$K, the CDW peak dynamics along the azimuthal direction are noticeably different (Fig. \ref{fig:timetraces}(f) and Supplementary Movie). For all measured times after $t_{\text{min}}$, an isotropic ring of intensity, distinct from the diffuse background, is visible at the CDW wavevector. Therefore, $C_0\approx1$ between $t_1$ and $t_\infty$. An azimuthal cut showing minimal intensity variation at the CDW wavevector is displayed in the inset of Fig.~\ref{fig:timetraces}(e). The diffuse CDW ring also exhibits persistent radial broadening out to $t=t_\infty$, in stark contrast to the recovery at 360K, which only exhibited radial broadening for a short time after $t_{min}$ (Fig. S6). The isotropic ring of intensity and radial broadening are signatures of a complete loss of both translational and orientational order of the CDW superlattice. These data are suggestive that the dislocation-type defects have dissociated into unbound disclination pairs. By circumventing the $1T$-to-$2H$ transition through photo-excitation, we are thus able to observe a liquid CDW state that is otherwise inaccessible in thermal equilibrium.

 \begin{figure*}[t]
    \centering	\includegraphics[width=1\columnwidth]{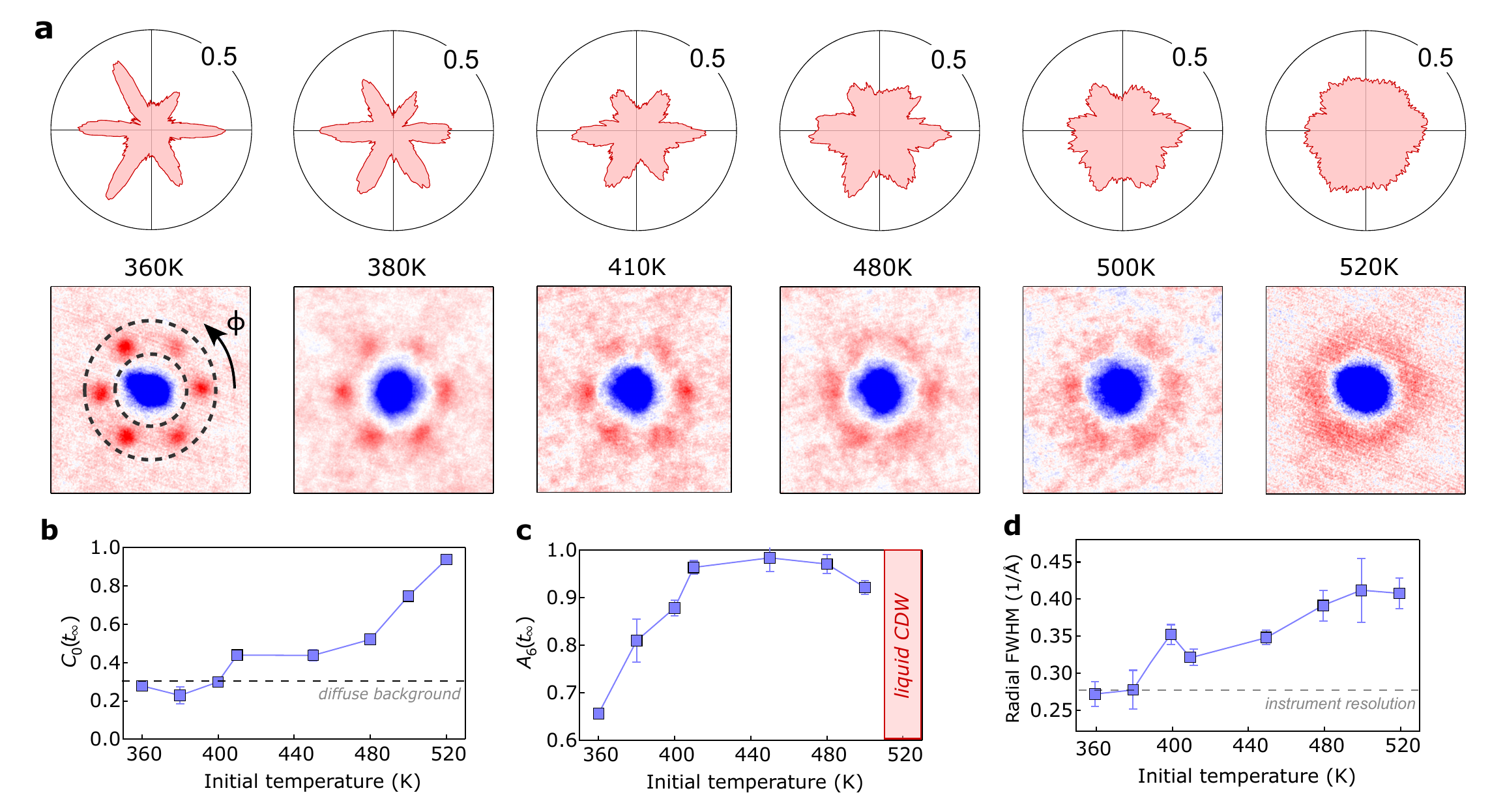}
    \caption{\textbf{Evolution of solid, hexatic, and liquid CDW states in the quasi-thermal regime.} \textbf{a}, Changes in diffraction intensity $I(t_{\infty})-I(t_{\text{min}})$ for various initial temperatures (bottom), and their respective azimuthal polar plots at the IC-CDW wavevector (top). \textbf{b}, Progression of the offset term $C_0(t_\infty)$ once the system has reached a quasi-equilibrium state, as a function of initial sample temperature. $C_0(t_\infty)$ of the diffuse background is plotted in gray for reference. \textbf{c}, Progression of the fundamental 6$\phi$ Fourier coefficient $A_6(t_\infty)$ as a function of initial temperature. \textbf{d}, Progression of the full-width-at-half-maximum (FWHM) of the IC-CDW peaks along the radial direction. The momentum resolution of our instrument is plotted in gray for reference.}
    \label{fig:heatingprogression}
\end{figure*}


Together, these data indicate that there are two distinct regimes of physical behavior. In the quench regime at short time scales ($t \lesssim t_2$), the system is still evolving and has yet to reach a thermalized state. Though relevant, thermal fluctuations in this regime do not dominate the behavior of the system. Conversely, in the quasi-thermal regime at longer timescales ($t \approx t_\infty \gg t_2$), the absorbed photons have effectively heated the illuminated region, and the system no longer changes with time. At the fluence used in these measurements, 6mJ/cm$^2$, the temperature at $t=t_\infty$ is raised by about 160K compared to the temperature before photo-excitation (Supplemental Note \RNum{5}). Therefore, for the data taken at initial temperatures of 360K and 520K, the final quasi-equilibrium temperatures, $T(t_\infty)$, are approximately 520K and 680K, respectively. Significantly, the latter final temperature is above both $T_{\text{CDW}}$ and the 1$T$-to-2$H$ transition temperature, which highlights how transiently heating the system maintains a superheated 1$T$ structure. Below, we show that there are two distinct phase transition pathways towards observing the CDW liquid -- a thermal and a non-thermal one.

\begin{figure}[t!]
    \centering	\includegraphics[width=1.0\columnwidth]{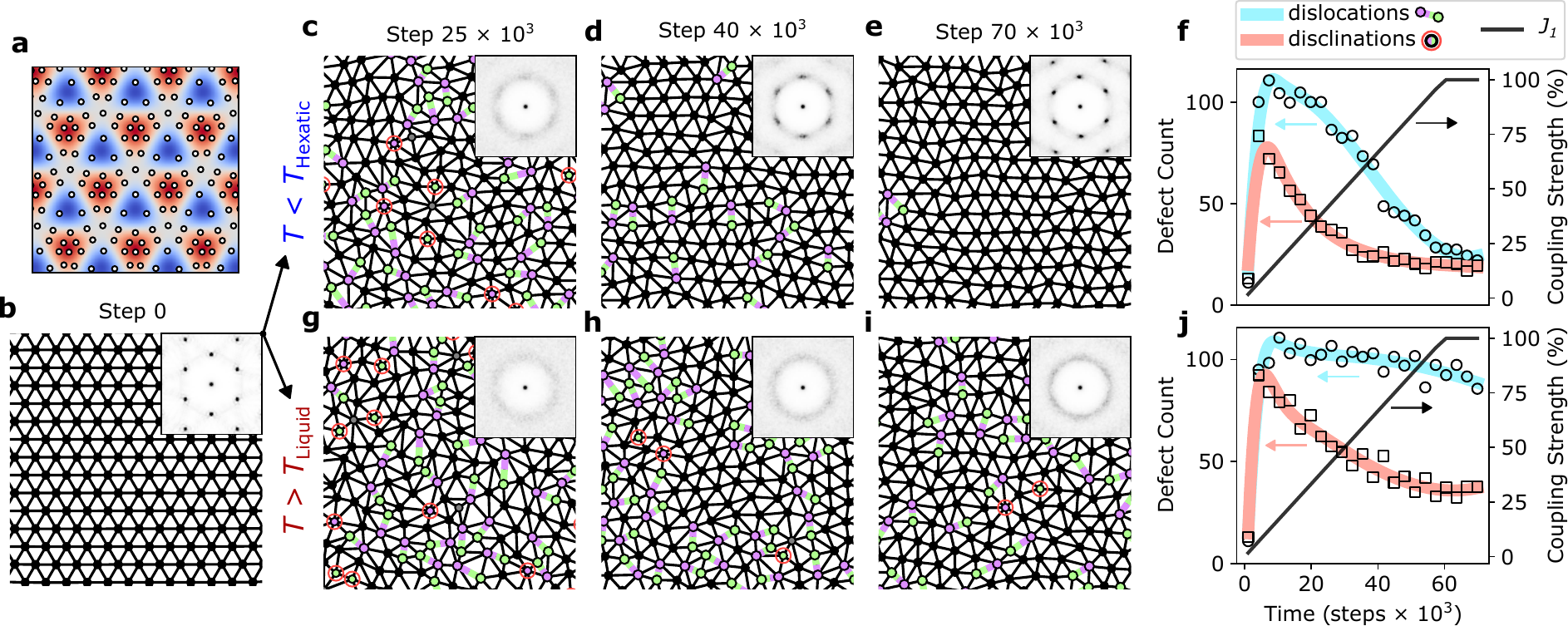}
    \caption{\textbf{Molecular dynamics simulation of a solid-liquid-hexatic transition with quenched interaction strength.} \textbf{a}, Schematic of IC-CDW with white filled points indicating tantalum atoms and the color map giving the coarse-grained charge density. \textbf{b}, Initial state of 2DMDS. \textbf{c-e}, Simulation snapshots at a temperature below the hexatic transition. The black lines show the Delaunay triangulation with defect sites in green (5 sites) and purple (7 sites). Disclinations are circled in red. The insets are simulated diffraction patterns. \textbf{f}, Number of dislocations (blue curve) and disclinations (red curve) as a function of time. The solid black line shows $J_1$ ramping back to $100\%$ strength. \textbf{g-j}, Simulation results at a temperature above the liquid transition.
    }
    \label{fig:2DMDS}
\end{figure}

We first show how the system evolves from the solid to liquid thermally at $t=t_\infty$ at various initial temperatures between $360$K and $520$K (Fig.~\ref{fig:heatingprogression}(a)). For initial temperatures within the range $380\text{K}<T<410\text{K}$, the system gradually progresses from the solid CDW state to the hexatic state, as $A_6(t_\infty)$ begins to increase and the peaks broaden radially (Fig. \ref{fig:heatingprogression}(c)-(d)). 
By 410K, $A_6(t_\infty)\approx1$, and dislocation pairs have unbound to destroy the quasi-long range translational symmetry of the CDW. Above 480K, $C_0(t_\infty)$ starts to increase significantly which is indicative of disclination pair unbinding, and at $520$K, the liquid CDW state becomes stabilized as $C_0(t_\infty)\approx1$. The radial peak width remains relatively constant after its initial rise near $T=410$K, consistent with the fact that both hexatic and liquid CDW states possess only short-range translational order. Taken together, the evolution of the CDW as a function of background temperature is reminiscent of that predicted by KTHNY theory. Although further experiments will be needed to verify whether the three dimensionality of the solid precludes the observation of the universal jumps predicted by the theory, the experimental phenomenology is nonetheless in good qualitative agreement with a two-step thermal phase transition through an intermediate hexatic state. Using an ultrafast heating protocol and transiently maintaining a superheated 1$T$ structure thus represents the first pathway to observing the CDW liquid.

The non-equilibrium dynamics in the quench regime ($t\lesssim t_2$) represent a second pathway towards the CDW liquid. To understand how the transition proceeds in the quench regime, we perform a two-dimensional molecular dynamics simulation (2DMDS) in the presence of two main forces between particles: (i) a repulsive, Yukawa-type force that is scaled by coupling constant $J_1$ and (ii) an orientational force that models the interaction with the background lattice.
The particles in the simulation represent the charge centers of the CDW, as illustrated in Fig.~\ref{fig:2DMDS}(a)-(b). Simulations are initialized in the ordered CDW state (Fig.~\ref{fig:2DMDS}(b)), and to model the quench, the Yukawa coupling, $J_1$, is set to zero and linearly ramped back to 100\% strength. $J_1$ is quenched in the simulation because its magnitude, which is proportional to the square of the order parameter, rapidly changes in the experiment as the CDW amplitude is suppressed to zero upon photo-excitation. The effect of temperature in the simulations is incorporated through a Langevin noise term acting on all particles with a variance proportional to the simulation temperature. A full description of the 2DMDS is provided in Supplementary Note \RNum{10}.

Snapshots of the simulations are shown for temperatures below the hexatic ordering temperature and above the liquid transition temperature in Fig.~\ref{fig:2DMDS}(c)-(e) and (g)-(i), respectively. For $T<T_\textrm{{Hexatic}}$, the system progresses from a solid to hexatic and back to solid as a function of time, which reproduces the experimental dynamics observed at 360K. For the higher temperature simulation ($T>T_{\textrm{Liquid}}$), the system remains liquid indefinitely following the quench, as is observed in the experiment at 520K. Though tuned by the coupling constant $J_1$ instead of temperature, the transition similarly proceeds though the proliferation of topological defects. Figures~\ref{fig:2DMDS}(f) and (j) show the number of dislocations and disclinations as a function of time (in terms of simulation steps) for $T<T_\textrm{{Hexatic}}$ and $T>T_\textrm{{Liquid}}$. In the former case, the defects eventually vanish at long times, while both disclinations and dislocations persist in the latter. These simulations highlight crucial role played by topological defects in the formation of the non-equilibrium CDW hexatic and liquid states through the quenching of a non-thermal parameter, which suggests that the quantum Kibble-Zurek mechanism may be operative in this regime~\cite{zurek2005dynamics, dziarmaga2005dynamics, polkovnikov2005universal}. Optical quenching thus presents a second pathway through which the CDW liquid can be observed.

Our work highlights how light pulses can be used to observe phases of matter that are hidden by intervening phase transitions. CDW re-crystallization in 1$T$-TaS$_2$ is revealed to proceed via liquid and hexatic CDW phases through two distinct pathways. Liquid CDWs, in particular, represent a new state of matter that may be prevalent, but have so far eluded detection, in a number of correlated electron systems. Whether the liquid CDW can be quenched into an amorphous glassy state, novel forms of liquid crystalline CDWs or other electronically textured phases remains to be seen, but present interesting avenues for further exploration. Ultimately, our work paves the way towards the discovery of exotic phases of electronic matter using non-equilibrium methods which can access regions of phase diagrams that may otherwise remain thermally forbidden.  

\nolinenumbers
\footnotesize
\vspace{-0.75em}
\section{Acknowledgements}
We thank S.E. Brown, R. Bruinsma, and X. Zhang for insightful discussions regarding this work. We thank M. Rasiah, S. Wang, J. Higgins, A. Ody, and A. Kulkarni for their instrumentation work in the keV UED set-up at UCLA. We acknowledge support from the US Department of Energy, Office of Science, Office of Basic Energy Sciences under award no. DE-SC0023017.

\section{Author contributions:  }
J.S.H.L. and T.M.S. performed the diffraction measurements. J.S.H.L. and T.M.S. prepared the samples for measurements. J.S.H.L. and T.M.S. built the keV UED beamline at UCLA with supervision from A.K. and P.M. G.K. grew the crystals for the experiment. J.S.H.L. performed the data analysis with theoretical input from T.M.S. and A.K. T.M.S. performed the molecular dynamics simulations. J.S.H.L. T.M.S. and A.K. wrote the paper with important input from all other authors. The work was supervised by A.K.
\vspace{-0.75em}

\section{Methods}

\subsection{Crystal growth and sample preparation}
\vspace{-1em}
Single crystals of $1T$-TaS$_2$ were grown in evacuated quartz ampoules (pressure less than 10 mTorr) in a three-zone gradient temperature furnace
(Lindberg/Blue-M) using iodine chemical vapor transport method \cite{di1973preparation,di1975effects}. The starting materials
were metal powder of Ta and Ti (purity better than 99.99\%, Alfa Aesar), sulfur powder (purity
99.999\% Puratronic), and iodine transport agent (purity 99.985\% Puratronic). The
stoichiometric ratio of metal and chalcogen powders (with slight excess of chalcogen) along
with traces of iodine were loaded in the ampoules and the growth proceeded in a gradient of
around 200 $^{\circ}$C/m at 700 $^{\circ}$C. The growth was terminated after 14 days with quenching the
samples from 700 deg to room temperature to avoid formation of interpolymorphic
transformations. The millimeter-size crystals have been characterized using structural (XRD),
spectroscopic (energy dispersive x-ray spectroscopy to verify stoichiometry) and low
temperature electrical transport probes.

Approximately 60 nm thick, 250 um x 250 um laterally sized flakes of $1T$-TaS$_2$ were sectioned along the a-b plane using an ultra-microtome fitted with a diamond blade. The sectioned flakes were transferred from water onto standard TEM copper mesh grids (90 $\mu$m hole widths). The grids are clamped onto our aluminum-based sample mounting apparatus. Silver-filled, ultra-high-vacuum compatible thermal paste was applied to the edges of the copper grids to improve thermal contact between the grid and the sample mount. 

\subsection{keV ultrafast electron diffraction with an external heating laser}
\vspace{-1em}
Experiments on the incommensurate-to-metallic transition in $1T$-TaS$_2$, both in equilibrium and through photoexcitation, were carried out at the University of California, Los Angeles (UCLA) using our table-top ultrafast electron diffraction (UED) apparatus. Photoexcitation (pump) pulses originate from a 1030 nm (1.20 eV), 180 fs Yb-based regenerative amplifier (RA) laser (PHAROS, Light Conversion) with tunable repetition rates up to 20 kHz. The pump pulses are directed into an optical parametric amplifier (OPA), allowing us to tune the wavelength of the output light pulses to 840 nm (1.48 eV). These pulses are subsequently focused to a 510 um x 510 um (FWHM) area on the sample. 

The electron pulses are generated by focusing the fourth-harmonic output (257.5 nm, 4.81 eV) of the 1030 nm light onto a 60 um x 60 um (FWHM) area on a poly-crystalline copper cathode. The photo-excited electron pulses are then accelerated to 40.4 kV in a d.c. field. To combat temporal-broadening due to space-charge effects, the pulses are sent through a radiofrequency (RF) cavity to achieve 375 fs (FWHM) temporal resolution~\cite{sutter2024ued}. The compressed electron pulses are focused by a magnetic lens to a 90 um x 90 um (FWHM) area on the sample. The diffracted electrons are collected and amplified by a micro-channel plate (MCP) detector with a phosphor screen (Photonis QS24215-1), and the final diffraction pattern is imaged using a CMOS sensor camera (FLIR Blackfly S USB3).  

In addition to the pump and probe pulses, we employ an additional heating laser to adjust the equilibrium temperature of our sample. The heating laser originates from the oscillator output of our Yb-KGW laser, with a wavelength of 1030 nm (1.20 eV) and at a repetition rate of 79.33 MHz. Considering the high repetition rate of the heating laser (79.33 MHz) compared with the repetition rate of our pump and probe pulses (1 to 4 kHz) and mismatch of their arrival times, we assume the heating laser pulses do not have an effect on the observed temporal dynamics after pump photoexcitation, providing only the effect of steady-state heating of the sample.  
 
\bibliographystyle{apsrev4-1}
\bibliography{refs}

\normalsize

\newpage

\setcounter{figure}{0}
\setcounter{table}{0}
\setcounter{equation}{0}
\setcounter{page}{1}
\renewcommand{\thefigure}{S\arabic{figure}}
\renewcommand{\theequation}{S\arabic{equation}}

\title{Supplementary Information for \texorpdfstring{\\ Topological phase transition to a hidden charge density wave liquid}{}}

\begin{titlepage}
\maketitle
\end{titlepage}

\newcommand{\pd}{\partial}
\newcommand{\beq}{\begin{equation}}
\newcommand{\eeq}{\end{equation}}
\newcommand{\bseq}{\begin{subequations}}
\newcommand{\eseq}{\end{subequations}}
\newcommand{\bpmat}{\begin{pmatrix}}
\newcommand{\epmat}{\end{pmatrix}}

\subsection{Determination of instrument momentum resolution}
\begin{figure}[hbt!]
	\centering
	\includegraphics[width=0.8\textwidth]{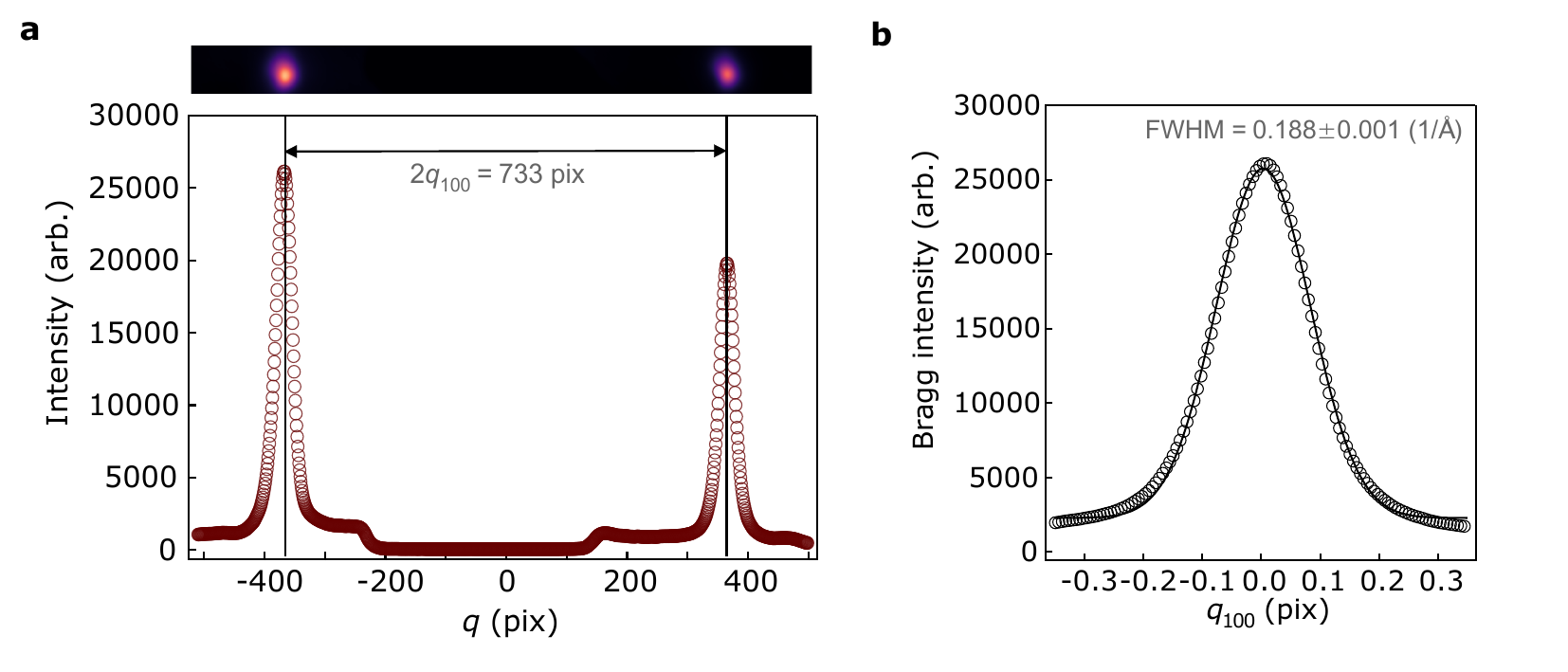} 

	\caption{\textbf{Measurement of the momentum resolution of our instrument.}
		\textbf{a}, Measurement of pixel separation between (100) and (-100) Bragg peaks. Considering the lattice constant of $1T$-TaS$_2$, a conversion from pixel length to inverse angstroms is obtained (1 pix = 0.00588 1/{\AA}). \textbf{b}, Measurement of the (100) Bragg peak FWHM ($0.188\pm 0.001$ 1/{\AA}), indicative of the momentum resolution of our instrument.}
	\label{fig:momentumRes} 
\end{figure}

The momentum resolution of our instrument is determined from a measurement of the width of the (100) Bragg peak. First, the camera pixel distance between the (100) and (-100) Bragg peaks is measured (Fig. \ref{fig:momentumRes}a). Then, considering the in-plane hexagonal lattice structure and the lattice constant $a$ of $1T$-TaS$_2$~\cite{yan2007anisotropic}, we use the equation $2q=8\pi/a\sqrt{3}$ to obtain our conversion factor between pixel distance to inverse angstroms ($1\text{ pix}=0.00588$ 1/{\AA}). Finally, the FWHM of the (100) Bragg peak is measured, and our momentum instrument resolution is determined ($0.188\pm 0.001$ 1/{\AA}) (Fig. \ref{fig:momentumRes}b).

\subsection{Static background intensity in IC-CDW diffraction peaks}

\begin{figure}[hbt!] 
	\centering
	\includegraphics[width=0.7\textwidth]{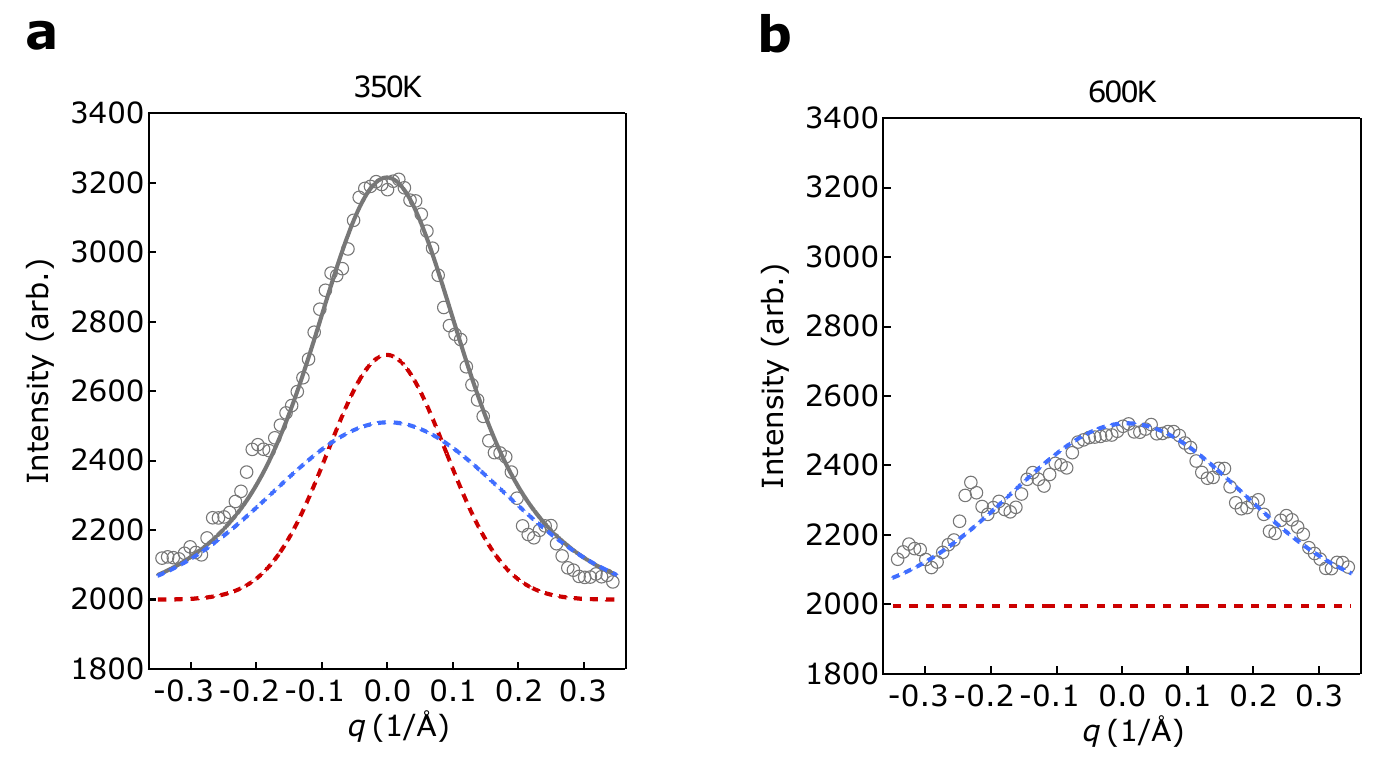} 

	\caption{\textbf{Static background subtraction via double-gaussian fitting.}
		\textbf{a}, Equilibrium diffraction profile of the $(-1-\delta,0,0)$ IC-CDW peak at a temperature of $T=350$K. The profile is fit with Eqn. \ref{eqn:doublegaus} (grey), and the individual Gaussian functions used for the fit are shown in red and blue for the IC-CDW peak and the background signal, respectively. \textbf{b}, Same as in \textbf{a}, except at a temperature of $T=600$K. Now, the IC-CDW peak has fully melted, and only the background signal remains.}
	\label{fig:defectgaussianfit} 
\end{figure}

A broad background signal is present in the equilibrium profile of the $(-1-\delta,0,0)$ IC-CDW peak; this signal developed over the course of about one month due to the application of many pump cycles on the sample. The signal presumably developed due to the gradual emergence of static, pump-induced defects in the crystal lattice. Its peak width extends broader than our instrument resolution, and the signal remains at a constant intensity level upon changes in temperature and also upon pump excitation. To extract the long-range ordered IC-CDW peak that is present above the background signal, we fit the overall intensity as the sum of two Gaussian functions:
\begin{equation}
    I(q)=I_0+I_1\text{ exp}\left(\frac{-q^2}{2\sigma_1^2} \right) +I_2\text{ exp}\left(\frac{-q^2}{2\sigma_2^2} \right) 
    \label{eqn:doublegaus}
\end{equation}
The fitted equilibrium profile of the $(-1-\delta,0,0)$ IC-CDW peak at temperatures $T=350$K and $T=600$K are shown in Fig. \ref{fig:defectgaussianfit}. Here, we see that the profile of the background signal $(I_2=510\pm10\text{, }\sigma_2=0.12\pm0.02$\AA$^{-1})$ remains constant through changes in the temperature of the sample. In all presented equilibrium data involving the background signal (Fig. 1), we subtract out the background signal to extract the true IC-CDW peak profile. To further clarify, such subtraction is not performed on the ultrafast diffraction data, where analysis is performed on differential intensity images (Fig. 2, Fig. 3).

\subsection{Translational long-range order of the IC-CDW state in equilibrium}

\begin{figure}[hbt!] 
	\centering
	\includegraphics[width=0.50\textwidth]{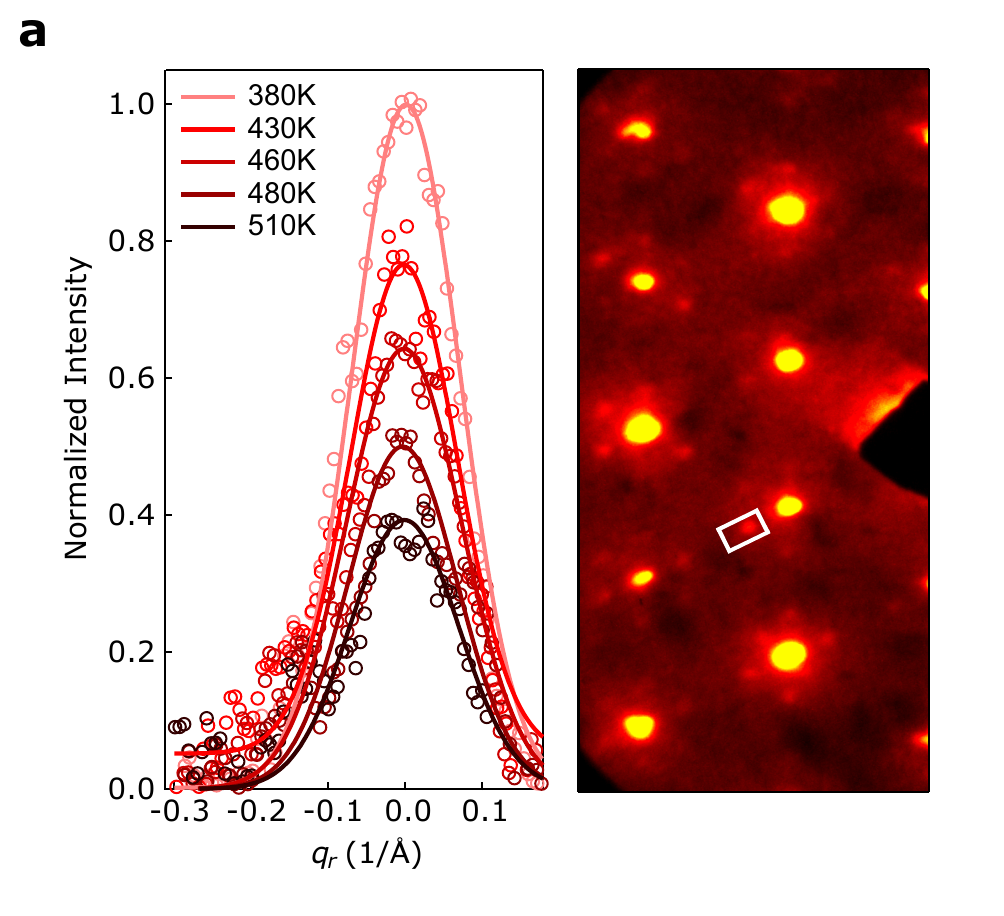} 

	\caption{\textbf{Translational long-range order during the IC-CDW-to-metallic transition.}
		\textbf{a}, Intensity profiles of the $(-1-\delta,0,0)$ IC-CDW peak along the radial direction, at selected initial temperatures. The specific line cut used is shown on the right. The IC-CDW radial peak width remains within the range of our instrument momentum resolution throughout the entire transition $(\text{FWHM}_{\text{radial}}=0.19\pm0.01$ \r{A}$^{-1})$.}
	\label{fig:radialpeakwidth} 
\end{figure}

As mentioned in the main text, when passing through the IC-CDW-to-metallic transition in equilibrium, broadening beyond our instrument momentum resolution is not detected in both the azimuthal and radial directions, implying that long-range order of the IC-CDW state is maintained in our investigated temperature range. Radial intensity profiles of the $(-1-\delta,0,0)$ IC-CDW peak at select temperatures are shown in Fig. \ref{fig:radialpeakwidth}, maintaining a FWHM of $0.19\pm0.01$ \AA$^{-1}$ throughout the transition.

\subsection{Determination of initial temperature before photoexcitation}
\begin{figure}
	\centering
	\includegraphics[width=0.8\textwidth]{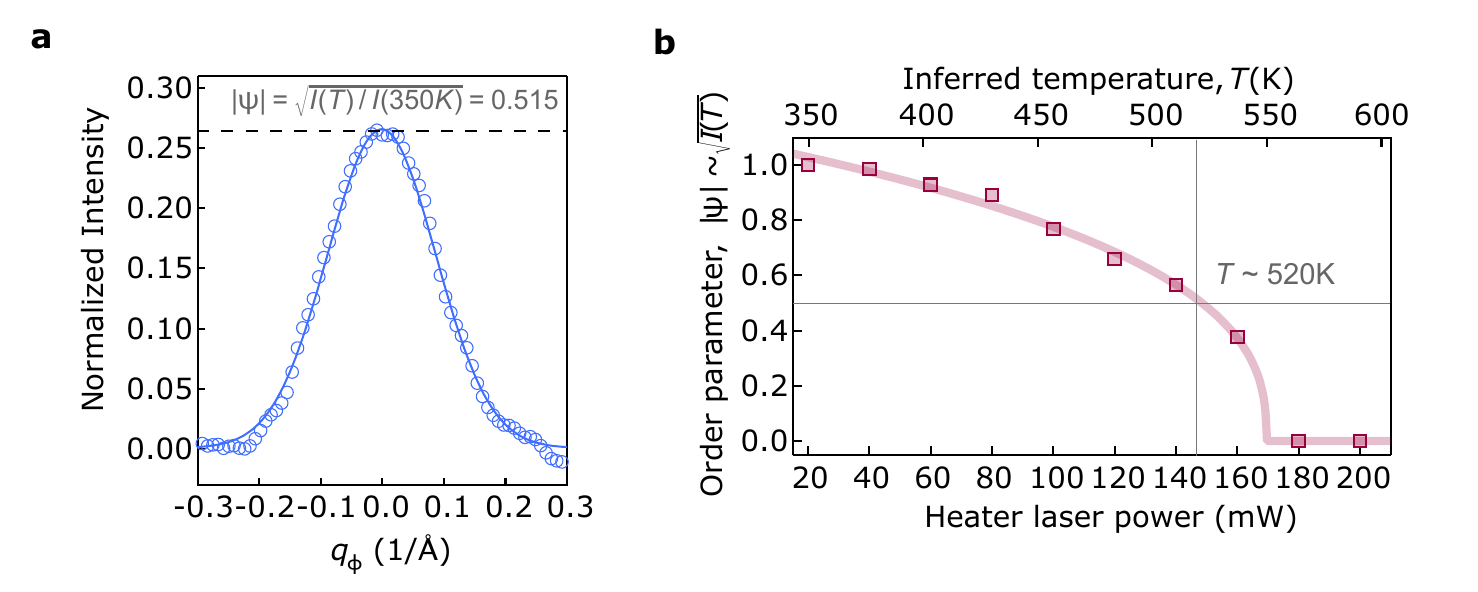} 

	\caption{\textbf{An example estimation of initial sample temperature before photoexcitation.}
		\textbf{a}, IC-CDW peak intensity profile before photoexcitation $(t=t_{<0})$, normalized to its intensity at 350K. Measurement of the order parameter is taken as the square root of the normalized intensity, $|\psi(T)| = \sqrt{I(T)/I(350\text{K})}$. \textbf{b}, After determining the initial value of the order parameter, it is compared against the equilibrium measurements of the IC-CDW-to-metallic phase transition. Here, we estimate $T= 520$K.}
	\label{fig:temp_example} 
\end{figure}

As mentioned in the main text, there are two sources of heating that raise the initial temperature of the sample: (1) steady-state heating of the pump pulses, and (2) steady-state heating of the external heating laser. To determine the initial temperature $T$, we begin by imaging the static diffraction pattern of the sample before pump excitation ($t=t_{<0}$). Then, we measure the intensity of the $(-1-\delta,0,0)$ IC-CDW diffraction peak as denoted in Fig. 1(b) and Fig. \ref{fig:radialpeakwidth}. We compare this peak intensity against that at 350K to determine the value of the order parameter at this temperature, i.e. we measure $|\psi(T)| = \sqrt{I(T)/I(350\text{K})}$. This value is compared against our equilibrium measurements of the order parameter (Fig. 1(d)), and a measurement of the initial temperature is obtained. Through this process, we estimate the uncertainty of all presented initial temperatures to be within a range of $\pm10$K. An example estimation procedure is shown in Fig. \ref{fig:temp_example}.

\subsection{Calculation of the lattice temperature after photoexcitation}
The temperature increase of our sample induced by an individual pump pulse can be estimated using the following equation:
\begin{equation}
    \Delta T_{\text{pump}} = 
    \frac{(1-R-T)f}{C_v\rho d}
\end{equation}
\noindent Here, $R=0.45$ is the reflectance of $1T$-TaS$_2$ at 840 nm \cite{li2019plane}, $f=6.0$ mJ/cm$^2$ is the pump fluence, $C_v = 420.3$ J/(kg K) is the specific heat of $1T$-TaS$_2$ \cite{manas2021quantum}, $\rho = 6.86$ g/cm$^3$ is the density of $1T$-TaS$_2$ \cite{haynes2016crc}, and $d = 60$ nm is the thickness of our sample. For the transmittance, we assume an exponential attenuation of the intensity of the pump pulse as it propagates through the sample, by the Beer-Lambert law:
\begin{equation}
    I(z)=I_0\exp(-\mu z)
\end{equation}
Here, $\mu=0.036 \text{ nm}^{-1}$ is the attenuation coefficient~\cite{li2019plane}. Using $z=60\text{ nm}$ for our sample, we obtain $T=0.10$. From these parameters, we estimate $\Delta T_{\text{pump}}\sim160$K.

\subsection{Averaging procedure of diffraction intensity profiles}

\begin{figure}[hbt!] 
	\centering
	\includegraphics[width=1.0\textwidth]{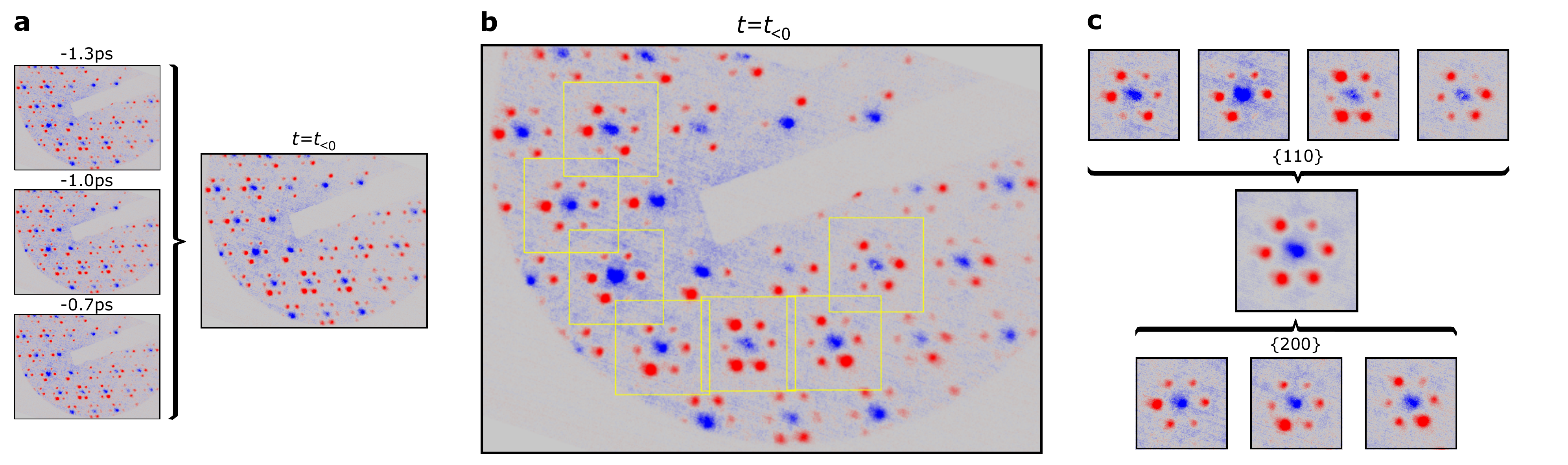} 

	\caption{\textbf{Example averaging procedure for enhanced signal-to-noise ratio.}
		\textbf{a}, Averaging several differential diffraction images near the selected time point $t_{<0}=-1\text{ ps}$. \textbf{b}, Selection of intensity regions-of-interests (yellow boxes) corresponding to Bragg peaks belonging to the visible \{110\} and \{200\} families of peaks. \textbf{c}, Averaging intensities from \textbf{b} together, to obtain our final difference image for $t=t_{<0}$ as shown in the main text.}
	\label{fig:averaging} 
\end{figure}

In the main text, we present differential diffraction profiles at selected time points with respect to $t=t_\text{min}$ (Fig. 2, Fig. 3). To enhance the signal-to-noise ratio, these difference images are obtained by the following procedure. First, several ($2\sim4$) differential intensity profiles $I(\textbf{q},t)-I(\textbf{q},t_\text{min})$ at times near the selected time points ($t_{<0}$, $t_1$, $t_2$, $t_{\infty}$) are averaged together. Then, we extract the profiles of all visible \{110\} and \{200\} families of peaks. Finally, these intensity profiles are averaged together to obtain the final images as shown in the main text. An example averaging procedure for ($T=360$K, $t=t_{<0}$) is shown in Fig. \ref{fig:averaging}.

\subsection{Dynamics of translational order after photoexcitation}
\begin{figure*}[hbt!]
	\centering
	\includegraphics[width=0.8\textwidth]{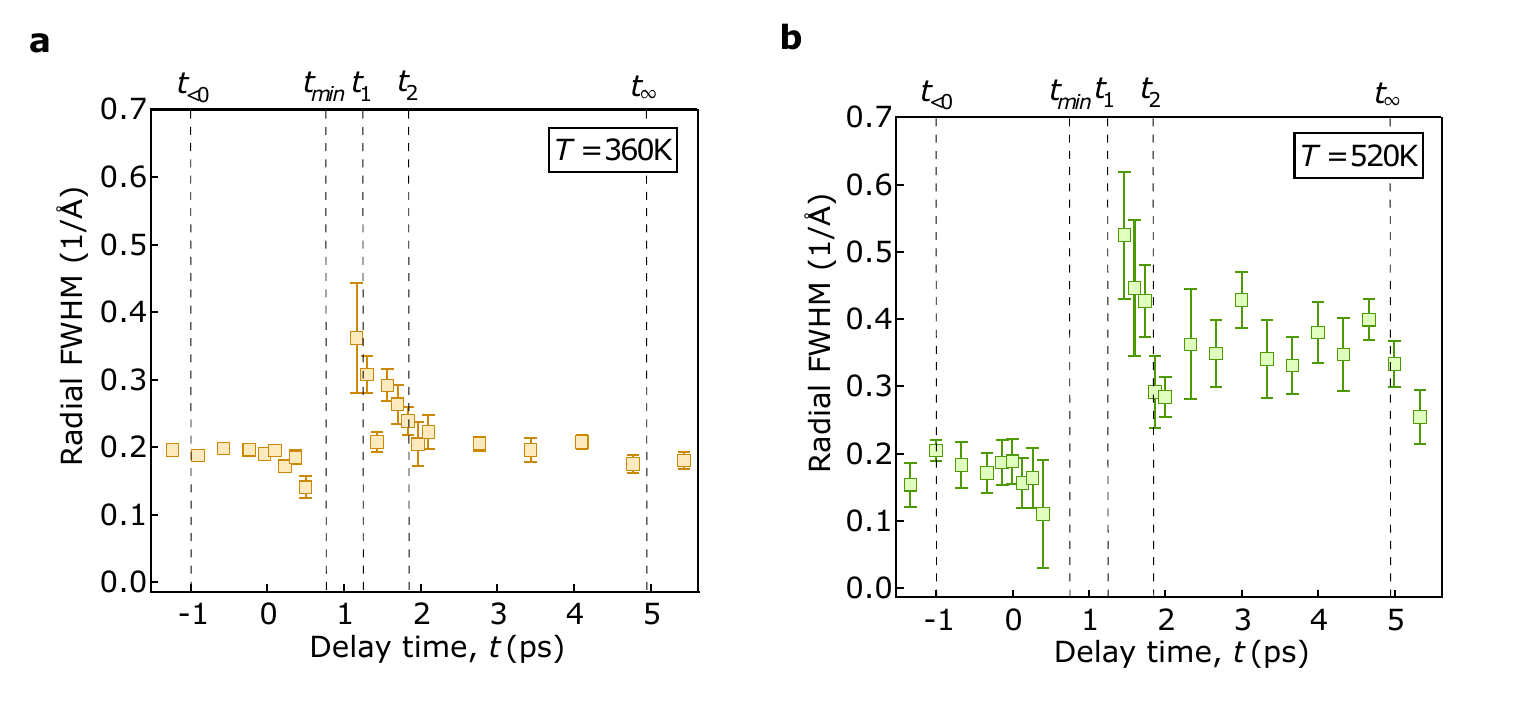} 

	\caption{\textbf{Dynamics of translational order of the IC-CDW superlattice after photoexcitation.}
		\textbf{a}, Dynamics of the FWHM of IC-CDW peaks along the radial direction, at an initial sample temperature of 360K. During the non-equilibrium re-crystalization of the IC-CDW state $(t_{\text{min}}<t<t_2)$, the radial peak width has increased beyond the instrument-limited value, indicating a transient loss of long-range translational order. At the quasi-equilibrium state $(t=t_\infty)$, the width has recovered to the instrument-limited value. \textbf{b}, Same measurement as shown in \textbf{a}, except now the initial sample temperature has been raised to 520K. After photoexcitation, the radial peak width has now increased beyond the instrument-limited value, even out to the thermalized state $t=t_\infty$.}
	\label{fig:radialfwhm} 
\end{figure*}

The dynamics of the translational order of the IC-CDW is inferred by measuring the changes in the radial width of the IC-CDW peaks after photoexcitation. The dynamics at 360K and 520K are shown in Figure \ref{fig:radialfwhm}. As mentioned in the main text, for $360$K, the radial peak width transiently broadens beyond our instrument resolution in the vicinity of $t=t_1$. This reflects the loss of long-range translational order in the IC-CDW superlattice, as expected for the hexatic IC-CDW state. At long times $t=t_\infty$, the radial peak width returns to the instrument-limited value; long-range translational order is restored as the system transitions back into the solid IC-CDW state. For $520$K, the radial peak width broadens near $t=t_1$, and persists out to $t=t_\infty$. This substantiates our observations of a liquid IC-CDW state that exists during both its non-equilibrium recovery time frame $(t_{\text{min}}<t<t_2)$ and after thermalization $(t=t_\infty)$.

\subsection{Phenomenological fitting procedure of intensity dynamics}
The following function is used to fit the dynamics of the averaged IC-CDW intensity after photoexcitation:
\begin{equation}
    f(t)=1-\frac{1}{2}\left(1+\operatorname{erf}\left(\frac{\left(t-t_{0}\right)}{\sigma\sqrt{2}}\right)\right)\cdot\left(I_{\infty}+I_{\text{min}}\exp\left(-\frac{\left(t-t_{0}\right)}{\tau}\right)\right)
\end{equation}
Here, $t_0$ is the time delay before arrival of the pump pulse, $I_{\text{min}}$ represents the minimum intensity value, $I_{\infty}$ is the value of the intensity at long time delay $(t=t_\infty)$, $\tau$ is the characteristic relaxation time of the recovery of the IC-CDW intensity, and $\sigma$ is the Gaussian root-mean-square width of the suppression of the IC-CDW intensity, limited by the temporal resolution of our instrument. 

For the dynamics of the Bragg peak intensity after photoexcitation, we use the following piecewise function:
\begin{equation}
    f(t)=
    \begin{cases} 
      1+\frac{I_{\text{peak}}}{2}\exp\left(-\frac{\left(t-t_{0}\right)}{\tau_\text{rec}}\right)\cdot\left(1+\operatorname{erf}\left(\frac{\left(t-t_{0}\right)}{\sigma\sqrt{2}}\right)\right) & t\leq t_0 \\
      1+\frac{I_{\text{peak}}}{2}\exp\left(-\frac{\left(t-t_{0}\right)}{\tau_\text{rec}}\right)\cdot\left(1+\operatorname{erf}\left(\frac{\left(t-t_{0}\right)}{\sigma\sqrt{2}}\right)\right)+I_{\text{DW}}\left(1-\exp\left(-\frac{\left(t-t_{0}\right)}{\tau_{\text{DW}}}\right)\right) & t\geq t_0
   \end{cases}
\end{equation}
Here, $t_0$ is the time delay before arrival of the pump pulse. $I_{\text{peak}}$ is the maximum intensity reached during the initial rise in Bragg intensity, $\sigma$ is the Gaussian root-mean-square width of the rise of the Bragg intensity, and $\tau_{\text{rec}}$ is the characteristic relaxation time of the subsequent drop in Bragg intensity. Both of these dynamics mirror that of the IC-CDW suppression and recovery, in accordance with the elastic scattering sum rule. $I_{\text{DW}}$ represents the value of the intensity at long time delay, and $\tau_{\text{DW}}$ is the characteristic relaxation time of the loss in Bragg intensity due to the Debye-Waller factor.

Finally, for the rise in the diffuse background, we use the following fit function:
\begin{equation}
    f(t)=\frac{I_{\infty}}{2}\left(1+\operatorname{erf}\left(\frac{\left(t-t_{0}\right)}{\sigma\sqrt{2}}\right)\right)
\end{equation}
Here, $I_\infty$ is the value of the diffuse intensity at long time delay, $t_0$ is the time delay before arrival of the pump pulse, and $\sigma$ is the Gaussian root-mean-square width of the rise in the diffuse intensity.

\subsection{Fourier decomposition of the IC-CDW intensity dynamics}
\begin{figure}[hbt!] 
	\centering
	\includegraphics[width=0.8\textwidth]{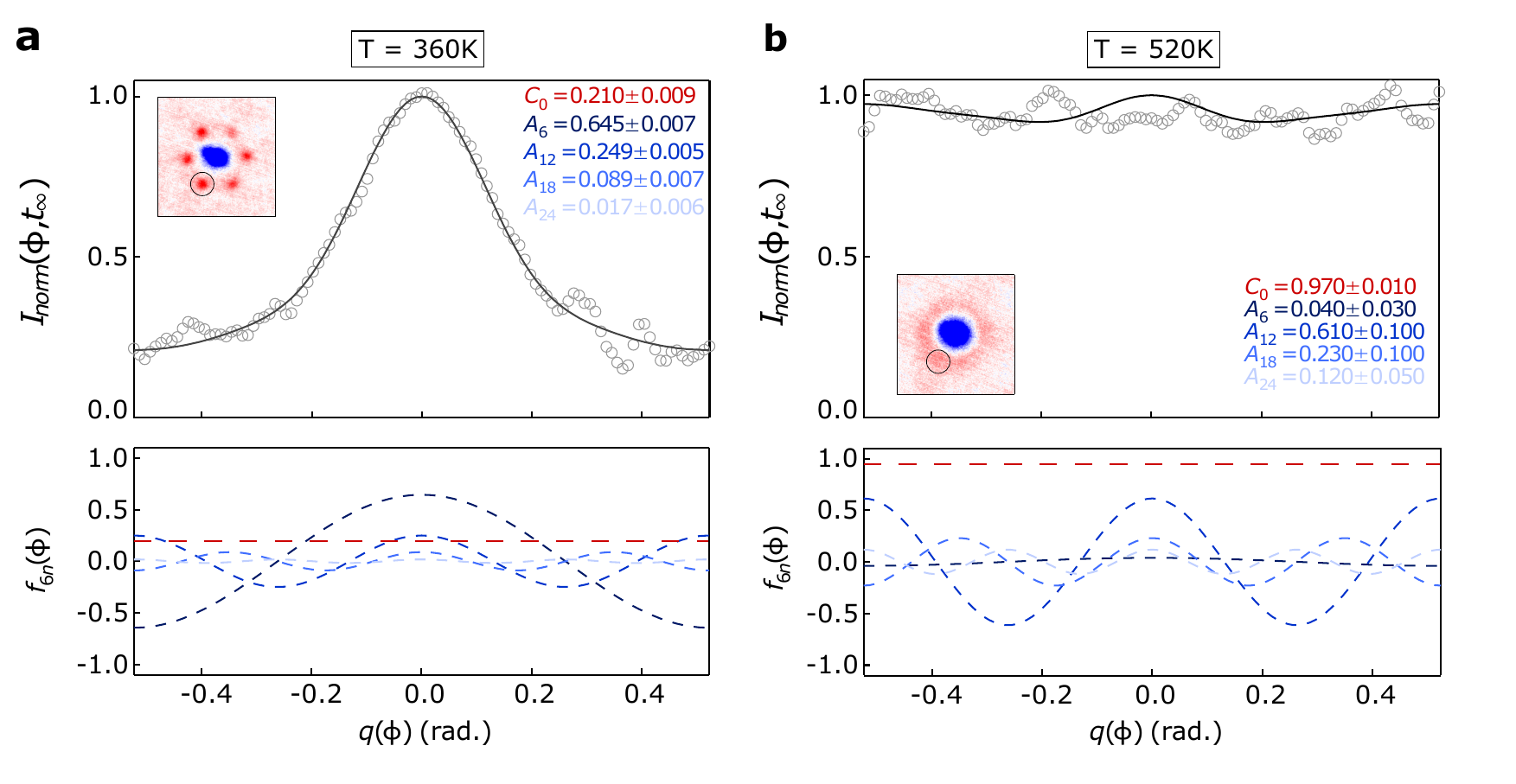} 

	\caption{\textbf{Example fitting procedure for extraction of Fourier component dynamics.}
		\textbf{a}, Azimuthal profile of the selected recovered IC-CDW intensity, as denoted in the inset, at long time delay $t=t_\infty$ and at an initial temperature of $T=360$K. The profile is fit with Eqn. \ref{eqn:fitfunc}, and the corresponding Fourier coefficients $A_{6n}$ are obtained (top). The individual harmonics comprising the fit function are plotted below, including the overall offset term. \textbf{b}, Same as in \textbf{a}, except at an initial temperature of $520$K. In this case, the azimuthal profile is dominated by the offset term $C_0$.}
	\label{fig:fitexample} 
\end{figure}

\begin{figure}[hbt!] 
	\centering
	\includegraphics[width=0.7\textwidth]{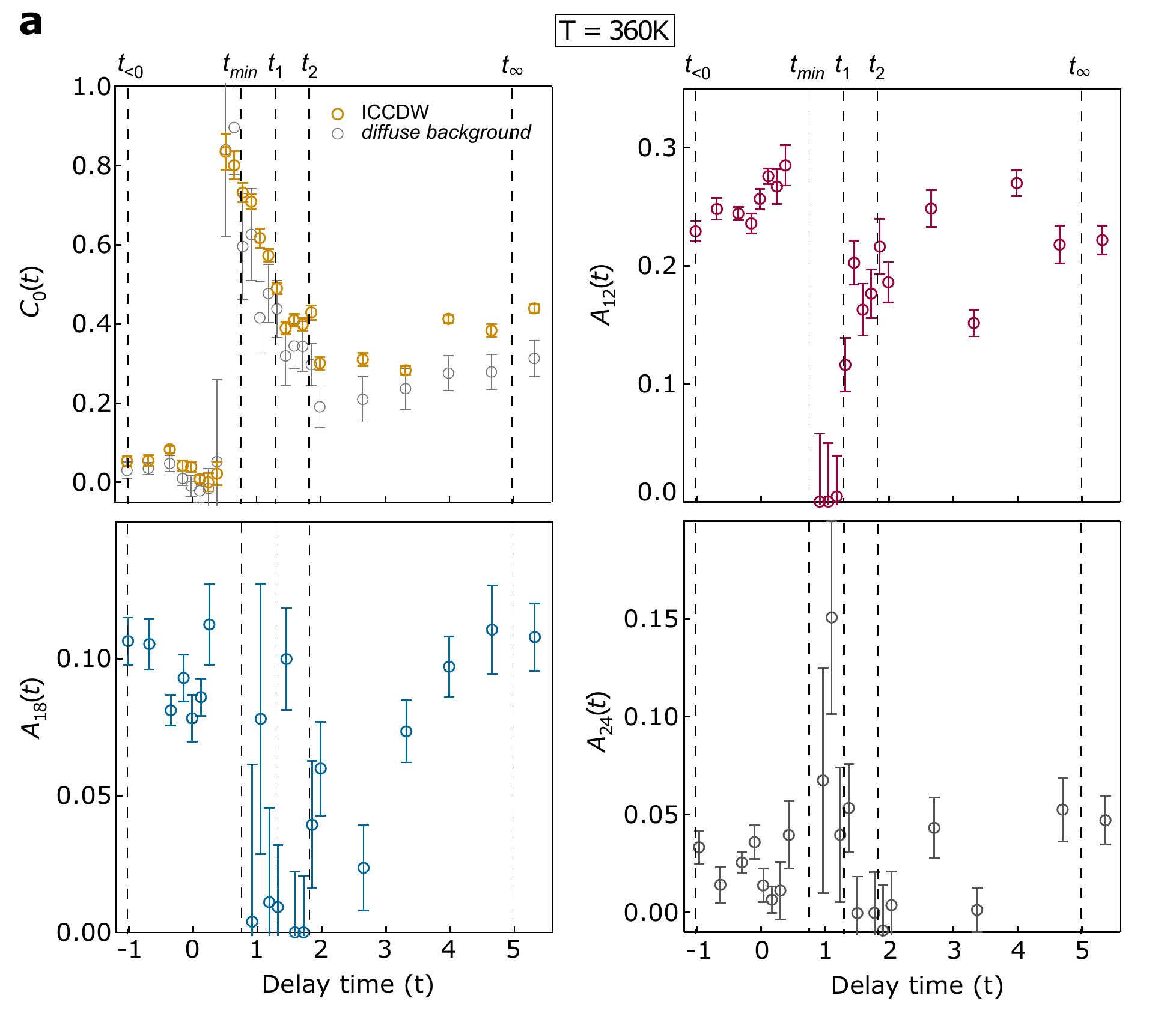} 

	\caption{\textbf{Low-temperature dynamics of all extracted Fourier coefficients.}
		\textbf{a}, Dynamics of $C_0,A_{12},A_{18},A_{24}$ Fourier coefficients after photo-excitation, at an initial temperature of $T=360$K. The IC-CDW peak selected for the fitting procedure is denoted in Fig. \ref{fig:fitexample}a. The $C_0$ dynamics of the thermal diffuse background is plotted alongside that of the IC-CDW peak for comparison.}
	\label{fig:highharmonics_lowtemp} 
\end{figure}

\begin{figure}[hbt!] 
	\centering
	\includegraphics[width=0.7\textwidth]{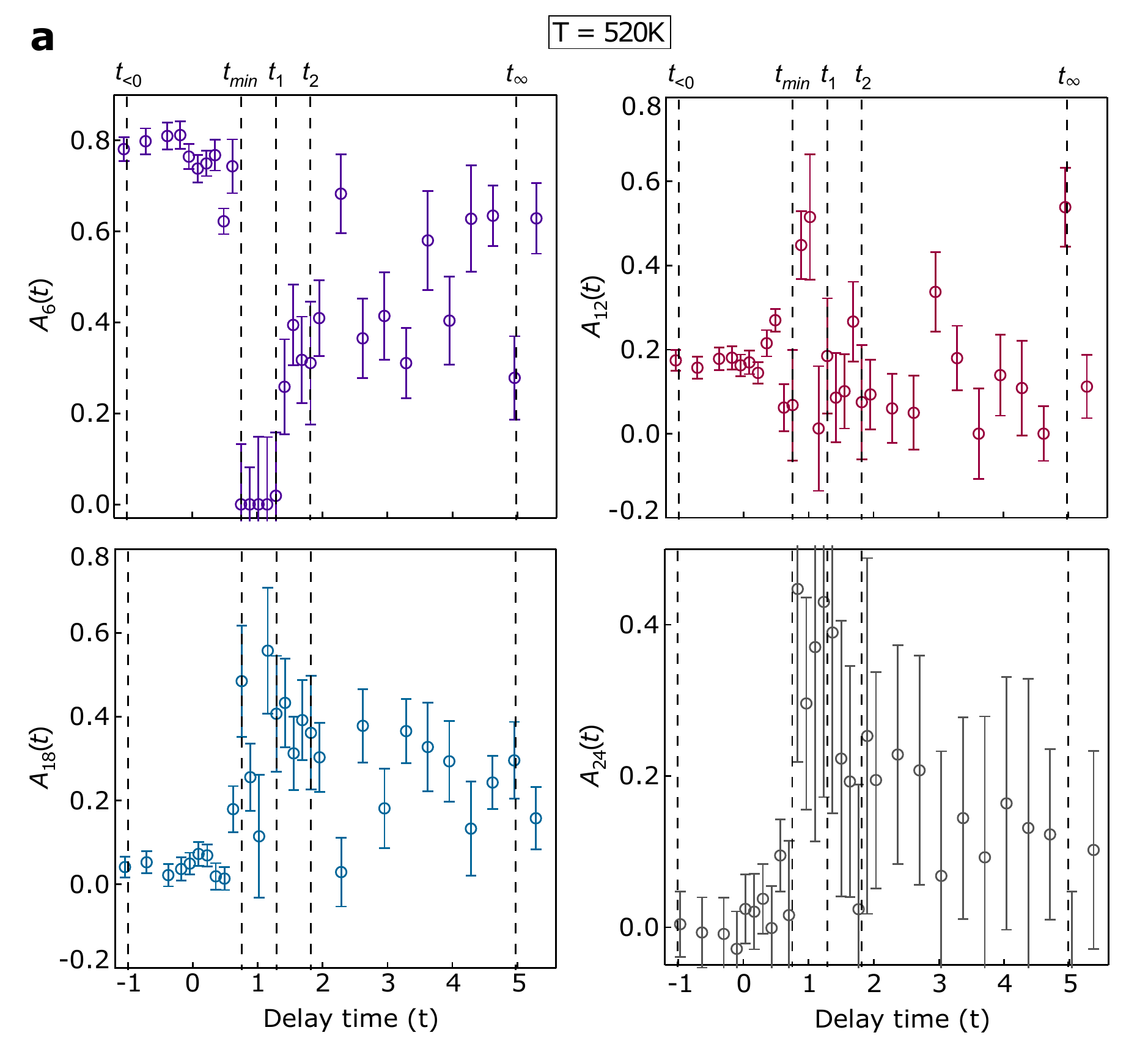} 

	\caption{\textbf{High-temperature dynamics of all extracted Fourier coefficients.}
		\textbf{a}, Dynamics of $A_6,A_{12},A_{18},A_{24}$ Fourier coefficients after photo-excitation, at an initial temperature of $T=520$K. The IC-CDW peak selected for the fitting procedure is denoted in Fig. \ref{fig:fitexample}b.}
	\label{fig:highharmonics_hightemp} 
\end{figure}

To quantify the dynamics of individual Fourier components of the azimuthal IC-CDW intensity profile after photoexcitation, we use the following procedure. First, at a given time point $t$, we consider the azimuthal intensity profile of the IC-CDW peak $I(q(\phi),t)$ as denoted in Fig. \ref{fig:fitexample}, spanning from $-\pi/6$ to $\pi/6$. The peak is normalized to an averaged maximum intensity value, taken in a region around the true maximum value:
\begin{equation}
    \phi_{\text{max}}=\text{argmax}[I(q_\phi,t)]
\end{equation}
\begin{equation}
     I_{\text{norm}}(q(\phi),t)=\frac{I(q(\phi),t)}{\frac{1}{2\phi_{\text{bin}}}\int_{\phi_{\text{max}}-\phi_{\text{bin}}}^{\phi_{\text{max}}+\phi_{\text{bin}}}\mathrm{d}\phi \text{ }I(q(\phi),t)}
\end{equation}
For all presented data, we use $\phi_{\text{bin}}=\pi/60$. This normalization procedure is done to decouple the dynamics of the functional form of the IC-CDW diffraction peak from its overall changes in intensity. After normalization, we fit the resulting diffraction peak with the first four partial sums of the Fourier series, normalized to 1 with an offset term $C_0$:
\begin{equation}
    f_{6n}(\phi)=A_{6n}\cos(6n\phi)
\end{equation}
\begin{equation}
    I_{\text{fit}}(\phi)=\frac{1-C_0}{2(A_6+A_{18})}\left(\sum\limits_{n=1}^4f_{6n}(\phi)-(1-2(A_6+A_{18}))\right)+C_0
    \label{eqn:fitfunc}
\end{equation}
Here, we constrain the coefficients $A_{6n}$ to be positive and such that $\sum\limits_{n=1}^4 A_{6n}=1$. An example of the fitting procedure and extracted coefficients is shown in Fig. \ref{fig:fitexample}, and dynamics of the coefficients not presented in the main text are shown in Fig. \ref{fig:highharmonics_lowtemp} and \ref{fig:highharmonics_hightemp} for the $T=360$K and $T=520$K measurements, respectively.

In particular, for the $360$K dataset, $C_0(t)$ for the IC-CDW peaks shows a sharp rise at $t=t_{\text{min}}$, and a subsequent relaxation to a quasi-equilibrium value at $t=t_2$. As mentioned in the main text, changes in $C_0(t)$ can arise from two sources: (i) disclination-type defects in the IC-CDW superlattice, and (ii) an increase in the diffuse background intensity. To rule out the former, we compare the dynamics of $C_0(t)$ between that for the IC-CDW peaks and for the diffuse background. The dynamics of the two are nearly identical between $t_{\text{min}}$ and $t_2$, but at long time delay, $C_0(t_\infty)$ for the IC-CDW peaks settles at a value slightly higher than that of the diffuse background, within error. This implies that there exists a small amount of residual disclination-type defects that has not yet recombined by $t=t_\infty$.

\subsection{Two-dimensional Molecular Dynamics Simulation}

To investigate the light induced solid-hexatic-liquid transition, we constructed a molecular dynamics simulation in which a collection of particles interact through a repulsive Yukawa potential. These individual particles represent the charge centers of the CDW. An angular interaction between the particle is introduced to simulate the bias from the underlying lattice to align the CDW with the axes of the crystal. The force on particle $i$ due to particle $j$ is,

\begin{align}
\vec{F}_{ij} = J_1 \vec{\nabla}\bigg[\frac{e^{-r_{ij}/\xi_1}}{r_{ij}}\bigg] + J_2 \sin(6\theta_{ij})e^{-r_{ij}/\xi_2} ~ \hat{t}_{ij}
\end{align}
where $r_{ij}$ is the distance between particle $i$ and particle $j$. The unit vector $\hat{t}_{ij}$ is orthogonal to $\hat{r}_{ij}$ - the unit vector pointing from $i$ to $j$. The orientation is such that $\hat{r}_{ij} \times \hat{t}_{ij} = +1$. This expression is controlled by four parameters: (the repulsive coupling: $J_1$, the angular coupling: $J_2$, the screening length of the repulsive interaction: $\xi_1$, and the screening length of the angular interaction: $\xi_2$). In addition to this force, each particle is acted upon by a Langevin noise term $\vec{\eta}(t)$. The noise term is randomly generated each step from a two-dimensional Gaussian distribution with a variance of $T/10$ - the simulation temperature. Also, each particle experiences a dissipative force of $\vec{F}^{(d)}_i = -\gamma \vec{v}_i$ where $\vec{v}_i$ is the velocity of particle $i$. This allows the system to settle into static structures. Finally, the particles are confined to a circular region about the origin of radius equal to 50 (in simulation length units).

The simulations were run in \texttt{Numba} accelerated \texttt{Python} \cite{numba}. Initially, 600 particles were generated with random positions. This system was time evolved through $10^5$ steps ($dt = 0.05$) at $T = 0.6$. The system settled into a nearly perfect 6-fold crystalline structure with an average interparticle distance of 3.97. This structure was then saved and used as the initial state for all later simulations. The system state is analyzed by constructing a Delaunay triangulation; defect sites with 5 or 7 neighbors are then easily identified. In addition, there are some anomalous sites with more than 7 or less than 5 neighbors. For dense clusters of defect sites, there is not an unambiguous sorting of the sites into clear dislocations and disclinations. Nevertheless, one can count the number of unbound sites (i.e. disclinations) by constructing a maximum cardinality matching on the bipartite graph of 5 sites and 7 sites. This is achieved using the Blossom Algorithm implementation of the \texttt{Networkx} library \cite{networkX}.

In the experiment, the IR pump melts the CDW order which results in a suppression of the repulsive coupling between charge centers. To model this, we performed simulations in which $J_1$ is initialized at zero and linearly ramped to 0.3 at a rate of $+5 \times 10^{-6}$ per step. Note that the coupling to the lattice is kept constant at $J_2 = 0.005$. The screening lengths are $\xi_1 = 2$ and $\xi_1 = 1$. The initial system state is perfect crystalline order; however, the lack of any repulsive interaction between the particles results in a rapid melting of this order. As $J_1$ is gradually increased, a liquid state is restored in which there is a peak in the pair-distribution-function without any orientational order. Depending on the temperature, further increase of $J_1$ restores the hexatic phase and finally the solid phase. At very low temperatures, the orientational coupling (parameterized by the constant $J_2$) is sufficient to prevent full melting of the initial crystalline order. In this case, a liquid state is never realized; the system goes from hexatic back to solid or simply remains solid.

These structures are assessed in the main text with simulated diffraction patterns that show the trend from sharp Bragg peaks to azimuthally elongated hexatic peaks to finally ring like liquid scattering. These diffraction patterns are constructed by summing the scattering signal from each particle. The diffraction intensity can be expressed as,
\begin{align}
I(\vec{q}) = \sum^N_n\exp(i ~{\vec{q} \cdot\vec{r}_n})
\end{align}
where the sum is over the $N=600$ particles and $\vec{r}_n$ is the position of the n$^{\text{th}}$ particle. To compute this as a discrete image, the free variable $\vec{q}$ is represented as a meshgrid $(q_{ij}^{(x)}, q_{ij}^{(y)})$ upon which $I_{ij}$ can be generated. Here $(i, j)$ denote the indices of the discrete diffraction image.
 
\end{document}